\documentclass[
 reprint,
 amsmath,amssymb, pra, aps
]{revtex4-1}

\usepackage{graphicx}
\usepackage{dcolumn}
\usepackage{bm}
\usepackage[separate-uncertainty=true, alsoload=synchem]{siunitx}
\usepackage[hidelinks]{hyperref}
\usepackage[utf8]{inputenc}

\usepackage{esvect}
\usepackage{pgf}
\usepackage{libertine}
\usepackage{libertinust1math}
\usepackage{blindtext}

\DeclareSIUnit{\arbitraryunit}{arb. u.}

\makeatletter
\DeclareRobustCommand{\AA}{%
  \leavevmode
  \vbox{\ialign{##\cr
    \hidewidth\char'27 \hidewidth\cr
    \noalign{\nointerlineskip\kern-1.4ex}
    A\cr
  }}%
}
\makeatother

\begin{document}
\preprint{APS/123-QED}

\title{
Very Large and Reversible Stark Shift Tuning of Single Emitters in Layered Hexagonal Boron Nitride
}
\author{Niko Nikolay$^{1,2}$}
\author{Noah Mendelson$^3$}
\author{Nikola Sadzak$^{1,2}$}
\author{Florian Böhm$^{1,2}$}
\author{Toan Trong Tran$^{3}$}
\author{Bernd Sontheimer$^{1,2}$}
\author{Igor Aharonovich$^3$}
\author{Oliver Benson$^{1,2}$}
\affiliation{
 $^1$ AG Nanooptik, Humboldt Universität zu Berlin,
 Newtonstraße 15, D-12489 Berlin, Germany \\
 $^2$ IRIS Adlershof, Humboldt Universität zu Berlin,
 Zum Großen Windkanal 6, 12489 Berlin, Germany\\
 $^3$ School of Mathematical and Physical Sciences, University of Technology Sydney, Ultimo, New South Wales 2007, Australia
}
\date{\today}
\begin{abstract}


Combining solid state single photon emitters (SPE) with nanophotonic platforms is a key goal in integrated quantum photonics. In order to realize functionality in potentially scalable elements, suitable SPEs have to be bright, stable, and widely tunable at room temperature. In this work we show that selected SPEs embedded in a few layer hexagonal boron nitride (hBN) meet these demands. In order to show the wide tunability of these SPEs we employ an AFM with a conductive tip to apply an electrostatic field to individual hBN emitters sandwiched between the tip and an indium tin oxide coated glass slide. A very large and reversible Stark shift of \SI{5.5(3)}{\nano\meter} at a zero field wavelength of \SI{670}{\nano\meter} was induced by applying just \SI{20}{\volt}, which exceeds the typical resonance linewidths of nanodielectric and even nanoplasmonic resonators. Our results are important to further understand the physical origin of SPEs in hBN as well as for practical quantum photonic applications where wide spectral tuning and on/off resonance switching are required.




\end{abstract}

\maketitle

Bright and tunable solid state single photon emitters (SPEs) are required for the realization of scalable quantum photonic technologies \cite{Awschalom2018, Atature2018}. Recently, SPEs in hexagonal boron nitride (hBN) have been extensively studied due to their promising optical properties. The hBN SPEs exhibit narrowband linewidths, fast excited state lifetimes, polarized emission and operate at room temperature \cite{Jungwirth2016, Tran2016a, Sontheimer2017, Kianinia2017, Mendelson2018}, which is attractive for many nanophotonics applications. The layered nature of hBN also offers potential advantages for integrating the SPEs with other 2D materials, to achieve hybrid quantum devices based on 2D systems \cite{Lethiec2014, Woessner2015}. Furthermore, the nanoscale hBN flakes can be coupled with foreign photonic resonators, such as waveguides, microdisks, or photonic crystal cavities, a crucial prerequisite for integrated nanophotonics systems \cite{Alem2009, Proscia2018}.

In order to exploit the functionality of an SPE-cavity system, tuning the SPEs' zero phonon line (ZPL) to a cavity's resonance is essential. First works on spectral tuning of hBN SPEs included strain or pressure tuning \cite{Xue2018}, as well as the application of an electric field by sandwiching the hBN flake between two graphene layers \cite{Grosso2017, Noh2018}. However, at room temperature or for plasmonic resonators switching into and out of resonance requires a reversible and wide-range tuning on the order of \SI{17}{\milli\electronvolt} \cite{Zhang2016, Chikkaraddy2016}. So far no SPE with such properties has been operated. Moreover, state of the art approaches to realize integrated elements for quantum nanophotonics often rely on the identification of pre-characterized SPEs and subsequent fabrication of photonic structures around it \cite{Shi2016a, Gschrey2013a, Dousse2008}. This requires a procedure to select individual SPEs from a larger ensemble.

In this work, we solve the critical issues mentioned above. We demonstrate the individually controlled and reversible tuning of SPEs in hBN using a high resolution conductive atomic force microscope (AFM) tip. Using this technique, high fields (up to \SI{500}{\mega\volt\per\meter}) can be applied to a nano flake of choice. The few nanometer thickness of hBN is ideal for this method, as the generated electric fields, which are perpendicular to the substrate, are ultimately limited by the distance between the AFM tip and the surface. Under our experimental conditions, we were able to achieve record dynamic tuning of over \SI{6}{\nano\meter} at room temperature.


\begin{figure*}[htb]
 \includegraphics[width=450pt]{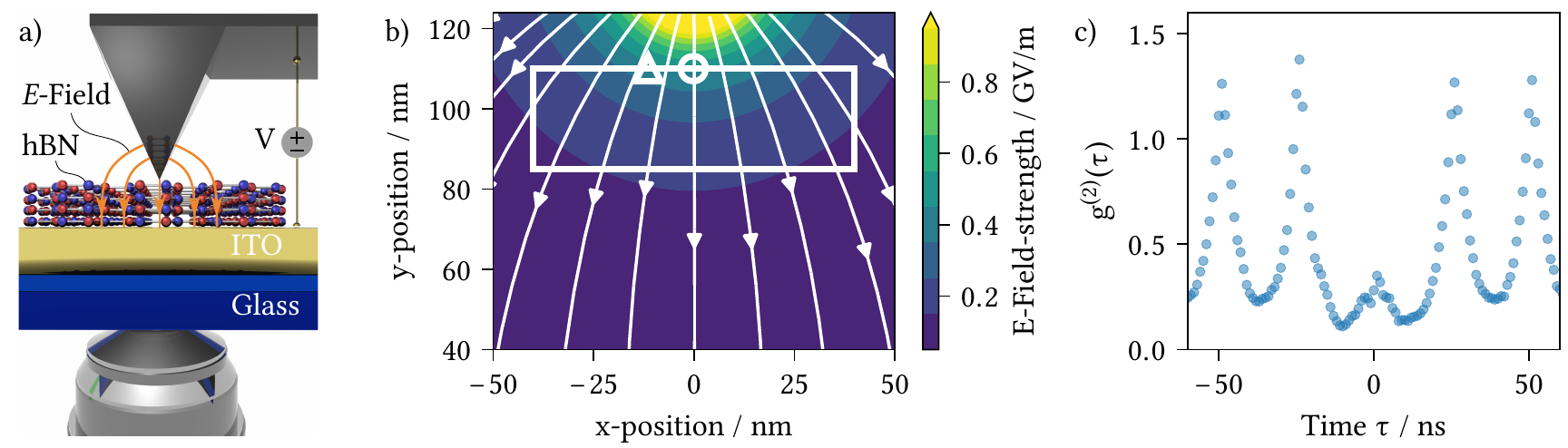}%
 \caption{\textit{Schematic representation of the experiment, electro static field distribution and $g^{(2)}$-function.} a) An hBN flake is located on an ITO-covered glass substrate. The oil immersion objective lens below excites the hBN SPE and collects its emission as an atomic force microscope tip can be used to deliver an electrostatic field causing a Stark shift of the ZPL. b) The electrostatic field strength between AFM tip (hBN starts at $y=\SI{125}{\nano\meter}$) and ITO (starts at $y=\SI{0}{\nano\meter}$) caused by the application of \SI{20}{\volt} is represented by the contour diagram. The superimposed stream flow chart shows the field orientation. The assumed SPE position (discussed in the text) is marked by the white rectangle. Points with the highest $|\vec\mu \vec E|$ and $|\vec E|^2$ within the assumed SPE area are marked by the triangle and the circle, respectively. c) The second order autocorrelation of the emitters fluorescence ($g^{(2)}(\tau)$) shows a clear antibunching at $\tau=0$ that indicates for primarily single photon emission.}
 \label{fig:schematics}
\end{figure*}

\begin{figure*}[htb]
 \centering
 \includegraphics[width=\textwidth]{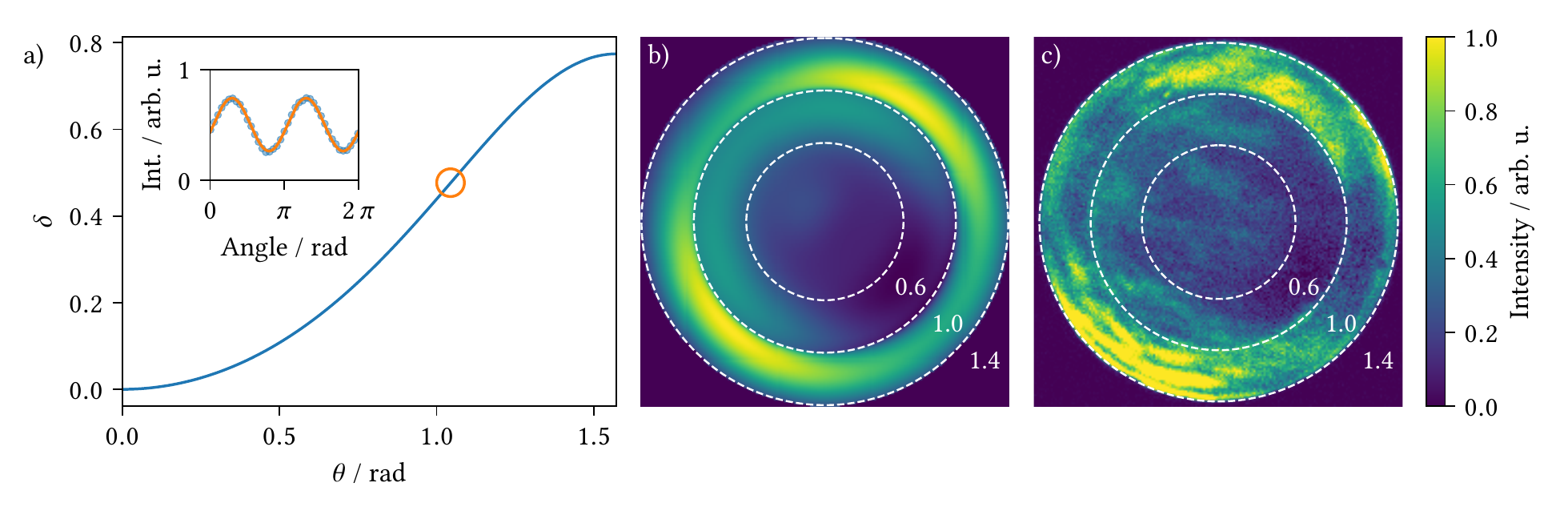}
 \caption{\textit{Determination of the dipole orientation.} a) Simulated degree of polarization $\delta$ with respect to the dipole out of plane angle $\theta$. The orange circle represents the measured $\delta$. In the inset, a corresponding polarization measurement of the fluorescence light (dots) and a fit (solid line) is shown. The signal was normalized to the total intensity detected by both APDs and corrected for its different detection efficiencies. b) Simulated Fourier image with a dipole orientation of $\theta = \SI{59.9(2)}{\degree}$ and $\phi = \SI{52.9(2)}{\degree}$ determined by the polarization measurement shown in a). c) Fourier image of the SPE taken with a $\mathrm{NA}=1.4$ objective lens. The striking similarity between b) and c) proves that the simulation is suited to derive the experimental results very well.
 }
 \label{fig:optics}
\end{figure*}

\begin{figure*}[htb]
 \includegraphics[width=510pt]{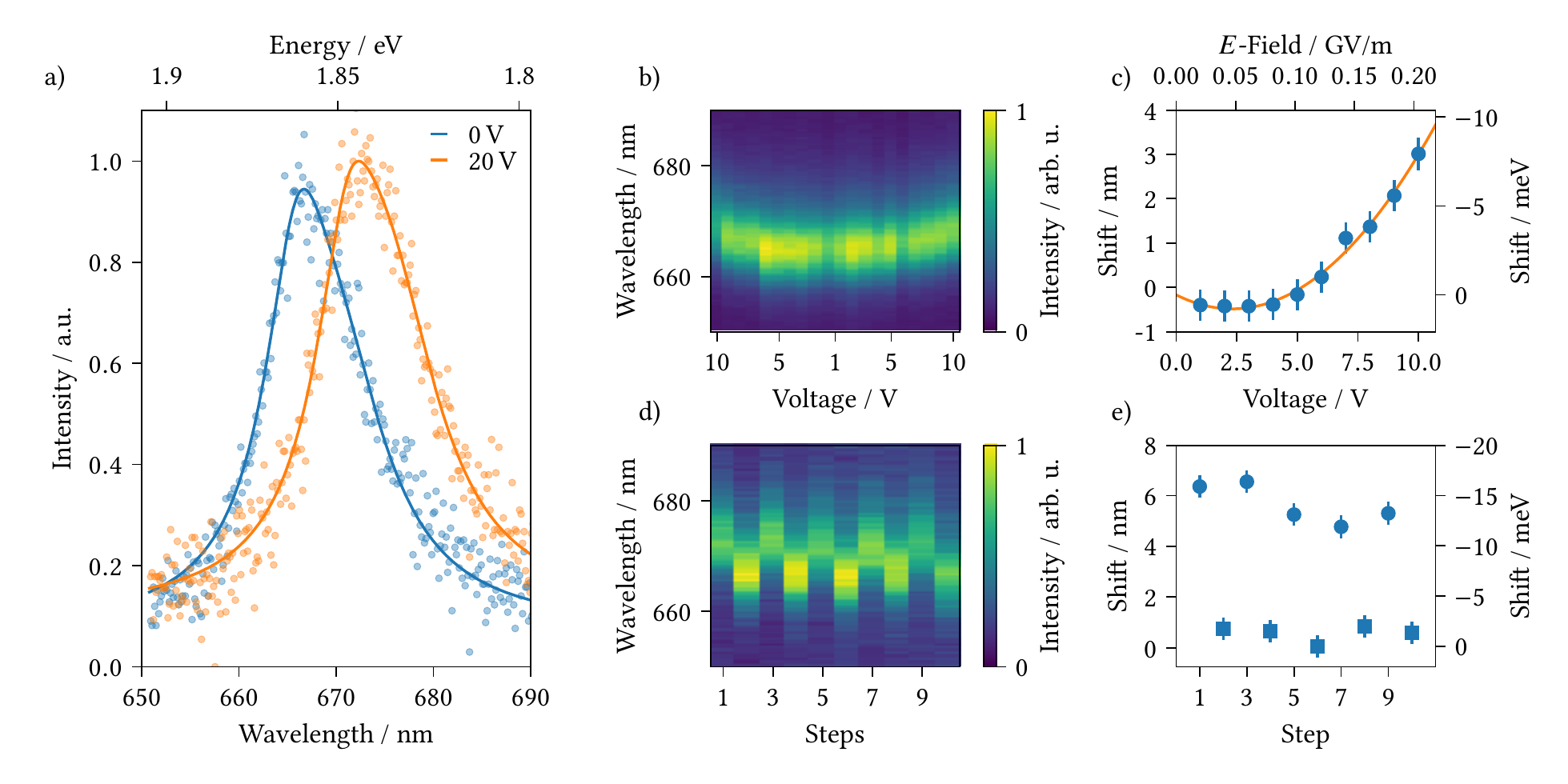}%
 \caption{\textit{Spectra of the Stark shifted hBN SPE fluorescence.} a) Measured spectra and corresponding fits of the unshifted emission in blue (left peak) and shifted in orange (right peak). This shift of \SI{5.9(6)}{\nano\meter} was recorded with a tip to substrate distance of $\approx\SI{125}{\nano\meter}$ and a Voltage of \SI{20}{\volt}. b) hBN-ZPL spectra recorded while the electric field strength between tip and ITO-layer was lowered and increased again (from left to right). c) Blue dots are spectral shifts determined by fits of \autoref{eq:fit} to the data shown in b). The solid line shows \autoref{eq:stark} that was fit to the data points. d) Spectra recorded while a voltage of \SI{20}{\volt} was switched on and off. e) Spectral shifts determined by fits to the data in d) show an average shift of \SI{5.5(3)}{\nano\meter} (\SI{15.4(8)}{\milli\electronvolt}).}
 \label{fig:starkShift}
\end{figure*}

In the following, the experimental setup and a pre-characterization will be introduced first, followed by a discussion about the measurement of the dipole orientation and the applied electrostatic field strength and direction. Together with a measurement of the $E$-field-dependent Stark shift, lower limits for dipole moment and polarizability are determined.


A schematic representation of the experimental setup is shown in \autoref{fig:schematics} a). An hBN flake hosting SPEs was sandwiched between a conductive AFM tip and a conductive indium tine oxide (ITO) coated cover slip. A Solea, PicoQuant laser with a central wavelength of \SI{540}{\nano\meter} (\SI{15}{\nano\meter} bandwidth) and a repetition rate of \SI{40}{\mega\hertz} was focused on the SPE from the substrate side via a high numerical aperture (1.4) oil immersion objective lens. SPE fluorescence was then collected by the same lens, passed through a \SI{610}{\nano\meter} long pass filter, an optional confocal pinhole, and finally guided either into a spectrometer, a Hanbury Brown and Twiss (HBT) interferometer, or an EMCCD camera recording a real or a Fourier image of the sample plane. A PL lifetime measurement (shown in the supplemental material) revealed an excited state lifetime of $\tau_\mathrm{hBN} = \SI{4.82(1)}{\nano\second}$, which is in the expected range for SPEs in hBN \cite{Tran2016}. Since the linear Stark shift depends on the alignment of the dipole moment of the emitter with respect to the electrostatic field direction, we must ensure that only the shift of the ZPL of a single emitter with known orientation is investigated. Thus, a second order correlation ($g^{(2)}$) function was calculated from photon arrival times recorded in the HBT setup. Blue dots in \autoref{fig:schematics} c) represent the $g^{(2)}$ function. The antibunching at $\tau=0$ is below $0.5$ and thus the emission can be considered as a predominantly single photon stream resulting from a SPE.


In order to identify the field strength and direction experienced by the SPE, the electrostatic field distribution present in the conducted experiment was simulated using COMSOL Multiphysics \textregistered. As parameters for the simulation an AFM tip radius of \SI{30}{\nano\meter}, a tip to ITO surface distance of \SI{125}{\nano\meter} (measured with the AFM), an hBN permittivity of 4 and a voltage of \SI{20}{\volt} were set. \autoref{fig:schematics} b) shows the field strength and direction indicated by the color coding and the arrows, respectively. The exact SPE position within the hBN flake is unknown, but two assumptions can be made to reduce the possible residence volume marked by the white rectangle. When the approached tip was scanned over the SPE, an intensity drop of up to \SI{35}{\percent} was observed. This drop is expected as the AFM tip alters the SPE radiation pattern, as well as the tip provides plasmonic decay channels potentially decreasing the external quantum efficiency. Simulations shown in the supplemental material map the decrease in intensity to a SPE depth below the hBN flake surface. The experimentally obtained drop is reproduced by the simulation at a minimal depth of \SI{15}{\nano\meter}. Furthermore, a noticeable Stark shift could only be seen when the lateral tip position was within an area of \SI{40}{\nano\meter} in diameter. At the edge of this area the electric field strength should be at least halved compared to its maximum in the center in order to quarter a quadratic Stark shift, rendering any shift in the experiment invisible by naked eye. This leads to a maximal depth of \SI{40}{\nano\meter}. Just the volume within those constraints (marked by the white rectangle in \autoref{fig:schematics} b) will be considered in the following discussions.


Next, we discuss the extraction of the dipole orientation from polarization dependent intensity measurements, which is crucial to determine the vectorial SPE dipole moment. Two angles, the in plane angle $\phi$ and the out of plane angle $\theta$ fully characterize its orientation. To determine both angles, a polarimetric measurement was performed \cite{Lethiec2014}. The horizontal and vertical polarization components of the SPE fluorescence are spatially separated using a polarizing beam splitter (PBS) and then individually detected via APDs at the corresponding output port of the beam splitter. A $\lambda/2$-plate before the PBS enables us to rotate the SPE polarization by the angle $\alpha$. To correct for intensity variations of the SPE during the measurement as well as for different detection efficiencies of each APD, we calculated the relative amount of the intensity detected by one APD as shown in the supplemental material, Eq. S3. The inset of \autoref{fig:optics} a) shows the resulting portion of the detected signal (dots), and a fit (line) of the following formula \cite{Lethiec2014}
\begin{equation}
    f(\alpha) = I_\mathrm{min}+(I_\mathrm{min}-I_\mathrm{max})\,\sin^2\left(\alpha+\phi\right),
\end{equation}
with the fit parameters $I_\mathrm{min}=\SI{0.262(3)}{\arbitraryunit}$, $I_\mathrm{max}=\SI{0.740(5)}{\arbitraryunit}$ and the in plane angle $\phi=\SI{52.9(2)}{\degree}$. The degree of polarization, given by $\delta = (I_\mathrm{max}-I_\mathrm{min})/(I_\mathrm{max}+I_\mathrm{min})$, is related to the out of plane angle $\theta$. We simulated a dipole with an orientation given by $\phi$ and $\theta$, located in an hBN flake (\SI{125}{\nano\meter} in diameter, measured with the AFM) on top of a glass cover slide with JCM wave, a 3D finite element Maxwell solver. From this simulation, we extracted the degree of polarization $\delta$ for any $\theta$ and compared it with the measured data, shown in \autoref{fig:optics} a), a detailed discussion can be found in the supplemental material. In this way we determined $\theta = \SI{59.9(2)}{\degree}$. To verify whether the simulated geometry is suitable to model the present experimental conditions, we quantitatively compare a simulated with a measured Fourier image, shown in \autoref{fig:optics} b) and \autoref{fig:optics} c) respectively. The dipole orientation in the simulation was given by the just determined angles $\phi$ and $\theta$, no free parameters were used. A clear similarity of both Fourier images can be seen.


To quantify relative spectral shifts, the ZPL central energy had to be determined. In order to account for the asymmetric nature of the ZPL at room temperature, we fitted a sum of two Lorentzian distributions given by
\begin{equation}
    \sigma = \frac{a_0\gamma_\mathrm{ZPL}^2}{(E-E_\mathrm{ZPL})^2+\gamma_\mathrm{ZPL}^2} + \frac{a_1\gamma_1^2}{(E-(E_\mathrm{ZPL}-E_1))^2+\gamma_1^2} + b
    \label{eq:fit}
\end{equation}
to the recorded spectra, with the amplitudes $a_0$ and $a_1$, the line widths $\gamma_\mathrm{ZPL}$ and $\gamma_1$, resonance energies $E_\mathrm{ZPL}$ and $E_1$, the offset $b$ and the measured spectral density $E$ (in $\SI{}{\electronvolt}$). All fits share the same values for $E_1$, $\gamma_1$ and $\gamma_\mathrm{ZPL}$. \autoref{fig:starkShift} a) shows photoluminescence (PL) spectra from the sandwiched SPE in few layer hBN (dots) and corresponding fits (solid lines) with \SI{0}{\volt} (blue) and \SI{20}{\volt} (orange) applied. A clear shift of \SI{5.9(6)}{\nano\meter} could be quantified in this way. 


We now proceed to study in detail the Stark shift behaviour and the modulation of the emission. \autoref{fig:starkShift} b) shows the ZPL spectra of a SPE as a function of the applied voltage. This experiment was done without readjusting the tip position, each spectrum was recorded for \SI{5}{\second}. First, the voltage was reduced, resulting in a blue shifted emission. Then, the voltage was increased back to its initial value, resulting in a red shift back to the original ZPL central energy, which indicates for a fully reversible shift. Fits of \autoref{eq:fit} to the averaged spectra (i.e. spectra taken at the same voltage were averaged) reveal central energies with respect to the applied voltage, shown in \autoref{fig:starkShift} c) as relative shifts. From this we can determine the dipole moment $\mu$ and the polarizability $\alpha$ by fitting the following formula adapted from Ref. \cite{Noh2018}:
\begin{equation}
    \Delta(\hbar\omega) = - |\vec\mu| |\vec E| \cos (\angle(\vec\mu,\vec E)) - \frac{1}{2} \alpha |\vec E|^2.
    \label{eq:stark}
\end{equation}
For this, the applied voltage must be related to an electric field seen by the SPE. As discussed before, the exact SPE position is unknown, and thus we are limited to estimating the minimum values for $|\vec\mu|$ and $\alpha$. Two points in the electrostatic vector field (represented in \autoref{fig:schematics} b)) were selected to relate the voltage to the electric field: one at which the scalar product $|\vec\mu \vec E|$ is maximum (marked by a triangle in \autoref{fig:schematics} b)) and one at which $|\vec E|^2$ is maximum (marked by a circle). The direction of $\vec\mu$ is given by the previously determined dipole orientation. At each point, a minimum value was determined by fitting \autoref{eq:stark} to the data points, where the x-axis was scaled for each maximum point according to the $E$ field simulation. In \autoref{fig:starkShift} c) the upper x-axis was exemplary scaled for the case of minimal $|\vec\mu|$, i.e. the triangle in \autoref{fig:schematics} b). The minimum values are given by $|\vec\mu|_\mathrm{Min} = (2.1 \pm 0.2)\,\mathrm{D}$ and $\alpha_\mathrm{Min} = (770 \pm 50)\,\AA^3$. They are in contrast to what was stated in the literature for hBN Stark tuning, where maximum values were given by $|\vec\mu|_\mathrm{Lit.} = 0.9\,\mathrm{D}$ and $\alpha_\mathrm{Lit.} = 150\,\AA^3$ \cite{Noh2018}. The discrepancy may result from the different dipole orientation with respect to the electric field, or the SPEs are of different atomic origin.

Finally, we demonstrate the reversibility of the shift and the stability of the emission over 10 cycles. For this purpose, we apply a square wave voltage between the AFM tip and the ITO-layer with an amplitude of \SI{20}{\volt}, a \SI{50}{\percent} duty cycle and a period time of \SI{5}{\second}. Note, that the AFM tip position was fixed during the whole measurement run. Again, the emission spectrum was recorded during this experiment, shown in \autoref{fig:starkShift} d). As before, fitting \autoref{eq:fit} to these spectra gives central positions shown in \autoref{fig:starkShift} e) as relatives shifts. A reversible shift of \SI{5.5(3)}{\nano\meter} (\SI{15.4(8)}{\milli\electronvolt}) in average was observed over 10 cycles. To the best of our knowledge, this is the largest reversible shift of any quantum system known to date.



In summary, we sandwiched an hBN flake hosting SPEs between a transparent conductive ITO layer and a conductive AFM tip. By applying a voltage between the two, a very large reversible Stark shift of \SI{5.5(3)}{\nano\meter} (\SI{15.4(8)}{\milli\electronvolt}) exceeding the resonance linewidth of typical nanodielectric and nanoplasmonic resonators \cite{Rybin2017, Zhang2016} was observed. Determining the SPE dipole orientation, its approximate position with respect to the AFM tip, and the electrostatic field distribution allowed us to translate the applied voltage into a vectorial electrostatic field experienced by the SPE. We found a linear and a quadratic Stark shift, described by the dipole moment of $|\vec\mu|_\mathrm{Min} = (2.1 \pm 0.2)\, \mathrm{D}$ and the polarizability of $\alpha_\mathrm{Min} = (770 \pm 50)\, \AA^3$. We could show that this very large Stark shift of the ZPL line is reversible and can be applied arbitrarily. This displays the potential to integrate selected SPEs in hBN in bisected plasmonic resonators, such as nanoparticle-on-metal plasmonic antennas \cite{DeNijs2015}. Such a configuration would represent a tunable plasmonic cavity quantum electrodynamical (CQED) system at room temperature.

Financial support from the German Ministry of Education and Research (BMBF) project "NANO-FILM", the Australian Research council (via DP180100077), the Asian Office of Aerospace Research and Development grant FA2386-17-1-4064, the Office of Naval Research Global under grant number N62909-18-1-2025 are gratefully acknowledged. I.A. is grateful for the Humboldt Foundation for their generous support. O.B. acknowledges the UTS Distinguished Visiting Scholars scheme.


\bibliography{references}

\end{document}


\title{
Very large and reversible Stark Shift tuning of selected single emitters in layered hexagonal boron nitride\\Supplemental material
}
\author{Niko Nikolay$^{1,2}$}
\author{Noah Mendelson$^3$}
\author{Florian Böhm$^{1,2}$}
\author{Toan Trong Tran$^{3}$}
\author{Bernd Sontheimer$^{1,2}$}
\author{Nikola Sadzak$^{1,2}$}
\author{Igor Aharonovich$^3$}
\author{Oliver Benson$^{1,2}$}
\affiliation{
 $^1$ AG Nanooptik, Humboldt Universität zu Berlin,
 Newtonstraße 15, D-12489 Berlin, Germany
}
\affiliation{
 $^2$ IRIS Adlershof, Humboldt Universität zu Berlin,
 Zum Großen Windkanal 6, 12489 Berlin, Germany
}
\affiliation{
 $^3$ School of Mathematical and Physical Sciences, University of Technology Sydney, Ultimo, New South Wales 2007, Australia
}
\date{\today}

\maketitle

\section{Lifetime measurement of the hBN single photon emitter}

\begin{figure}[tbp]
 \includegraphics[width=255pt]{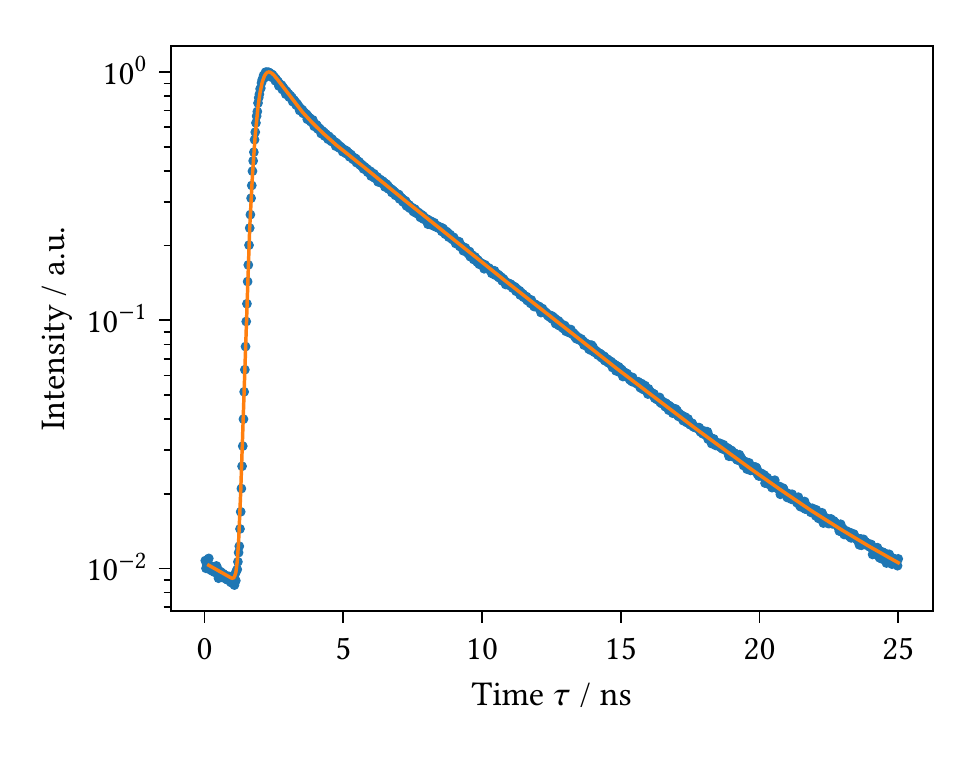}%
 \caption{\textit{Lifetime measurement fit.} Dots represent the experimental data of a lifetime measurement. The solid line shown the fit of a bi-exponential decay convoluted with an instrument response function, giving the lifetime $\tau_\mathrm{hBN} = \SI{4.82(1)}{\nano\second}$.}
 \label{fig:lifetime}
\end{figure}

To measure the excited state lifetime of the hBN SPE, we recorded time differences between the applied laser pulse and the photon arrival at on of the APDs. A resulting normalized histogram is shown in \autoref{fig:lifetime}, represented by dots. The solid line is a fit to the data points given by a circular convolution (denoted by $*$) of a double exponential decay with the instrument response function:

\begin{align}
    f(\tau) &= IRF * LT \\
    LT &= a_1 \mathrm{e}^{-t/\tau_1} + a_2 \mathrm{e}^{-t/\tau_2} + b.
\end{align}

With the amplitudes $a_i$, lifetimes $\tau_i$ and offset $b$ fit parameters (with $i = 1,2$). 

\section{Effect of relative position of AFM tip on collected intensity}

\begin{figure}[bp]
 \includegraphics[width=468pt]{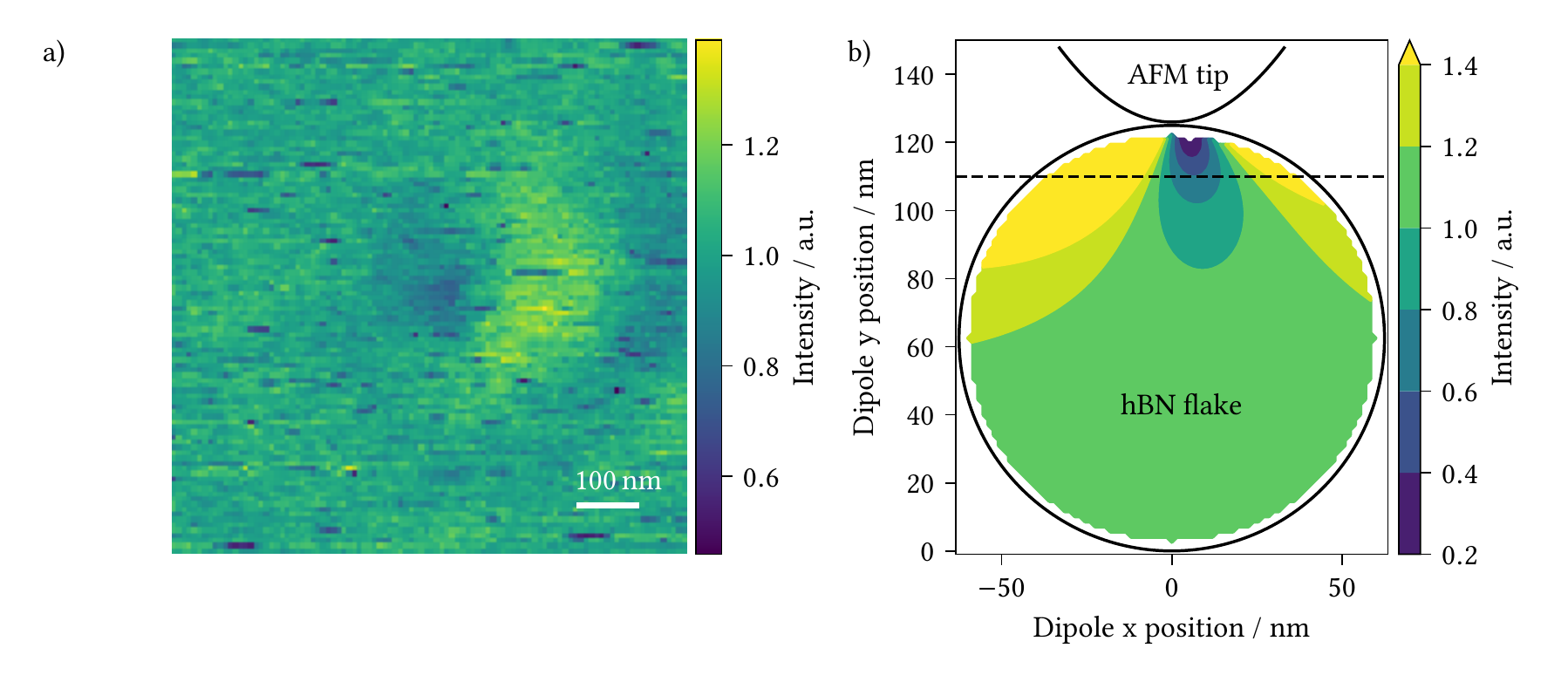}%
 \caption{\textit{Distance dependent intensity drop.} a) Intensity map of the SPE with respect to the AFM tip position, normalized to the intensity recorded without the tip attached. A decrease to 0.65 to 0.8 was observed. b) Simulated decrease in intensity when the AFM tip is pressed on the emitter. The circle represents the boundary of the hBN flake, the dotted line mark the SPE depth of \SI{15}{\nano\meter}.}
 \label{fig:intdrop}
\end{figure}

As discussed in the main text, when the AFM tip approaches the sample, it alters the SPE radiation pattern and provides plasmonic decay channels. This results in a decrease of the detected emission intensity, as shown in \autoref{fig:intdrop} a). Here the intensity of the SPE emission was recorded with respect to the AFM tip position while scanning over the hBN flake. We assume that the largest intensity loss of \SI{35}{\percent} occurs when the tip was closest to the SPE. As a simple model, we further assumed that in this situation the AFM tip was placed directly at the top of the hBN flake. For this case, we simulated a radiating dipole placed at different x/z positions within an hBN flake with a fixed dipole orientation, which was determined experimentally. The intensity detected by an NA 1.4 objective lens in the presence of an AFM tip, normalized to the intensity detected without the tip with respect to the dipole position, gives the heat map shown in \autoref{fig:intdrop} b). If the dipole is far from the tip (which is placed on top of the drawn circle), the intensity ratio approaches one, as expected. However, if the dipole position approaches the tip depending on the horizontal displacement, either an increase or a decrease in intensity can be seen. An intensity below 0.6 can be seen at a depth of \SI{15}{\nano\meter}, marked by the dotted line.

\section{Normalization of the measured polarization intensity}

The intensity recorded at each output port of the polarizing beam splitter is denoted by $I_\mathrm{APD_1}$ and $I_\mathrm{APD_2}$. To correct for overall intensity variations of the SPE during the measurement we calculated the relative intensity detected at one port by normalization to the intensities detected on both ports. In order to correct the different detection efficiencies of the individual APDs, we have additionally normalized the relative signal to its value averaged over $2\pi$ (denoted by the horizontal line):

\begin{equation}
    I_\mathrm{Norm}(\alpha) = \frac{1}{2}\,\overline{\frac{I_\mathrm{APD_1}}{I_\mathrm{APD_1}+I_\mathrm{APD_2}}}^{-1} \left(\frac{I_\mathrm{APD_1}}{I_\mathrm{APD_1}+I_\mathrm{APD_2}}\right).
\end{equation}

\section{Sample preparation}

The sample was prepared by first spin coating a colloidal solution of hBN-flakes (from Graphene Supermarket) onto a sacrificial Si substrate, which was then annealed at \SI{850}{\celsius} for 1 hour in a \SI{1}{\milli\torr} argon atmosphere. PMMA (A5) was then coated on the sample, baked for \SI{5}{\minute} at \SI{90}{\celsius}, before the PMMA/hBN film were freed from the substrate by etching in \SI{1}{\Molar} KOH solution. After washing 3 times in milli-Q water, the sample was then transferred to a pre-prepared ITO coated cover slip, heated to \SI{120}{\celsius} to promote adhesion between the hBN and ITO slide, before the PMMA film was washed away by dissolving in warm acetone overnight, analogous to Ref. \cite{Tran2017}. 



\bibliography{references}